\documentclass{rmaa}
\usepackage{amsmath}
\usepackage{stackengine}
\usepackage{xcolor}

\definecolor{rkka}{RGB}{219,66,32}

\newcommand{\degr}{^\circ\,}

\newcommand{\eb}{\begin{equation}}
\newcommand{\ee}{\end{equation}}
\newcommand{\norm}[1]{\left\lVert#1\right\rVert}

\title{Mass ratios of long-period binary stars resolved in precision astrometry catalogs of two epochs}

\author{Valeri V. Makarov
\affil{U.S. Naval Observatory, 3450 Massachusetts Ave., Washington, DC 20392-5420, USA}
}

\resumen{}
\abstract{
Mass ratios of widely separated, long-period, resolved binary stars can be directly estimated from the available data
in major space astrometry catalogs. The method is based on the universal principle of inertial motion of the system's
center of mass in the absence of external forces, and is independent of any assumptions about the physical parameters
or stellar models. Using two separate astrometric epochs, such as the ESA's
Hipparcos and Gaia mission results, the differences in components' positions and proper motions
generate sufficient condition equations to eliminate the unknown motion of the barycenter and compute the mass ratio
with possible redundancy. The application is limited by the precision of input astrometric data, the orbital period
and distance to the system, and possible presence of other attractors in the vicinity, such as in triple systems.
A generalization of this technique to triples is proposed, as well as approaches to estimation of uncertainties.
The known long-period binary HIP 473 AB is discussed as an application example, for which a $m_2/m_1=0.996^{+0.026}_{-0.026}$
is obtained.
}

\shorttitle{Mass ratios of binaries from two-epoch astrometry}
\shortauthor{Valeri V. Makarov}

\begin{document}
\maketitle

\section{Introduction}
\label{sec:intro}

The majority of stars in the Galaxy have masses less than the solar mass. Almost half of the regular solar type
field stars are
also components of binary or multiple gravitationally bound systems \citep{tok14}.
A binary star's orbit and ephemerides are described by seven orbital parameters (elements). From the astrometric point of
view, an additional five parameters representing the position at a given epoch, proper motion, and parallax of the barycenter
are required to completely characterize the absolute location of the components on the celestial sphere at a given time. 
The most important
elements for astrophysical investigations are the size (semimajor axis), the period, and the eccentricity. The former two
yield the total mass of the binary via Kepler's third law. But the distribution of masses between the components often remains
unknown, because the barycenter is not directly available from astrometric observations. 
In some rare cases, the mass ratio can be
directly inferred from spectroscopic radial velocity measurements for double-lined spectroscopic binaries (SB2) of similar
brightness, which
limits this method to mostly pairs of similar masses. When a good astrometric orbit, which is differential for resolved
pairs, is also available for
an SB2, accurate individual component masses can be obtained, providing valuable and still rare data for astrophysics.
Furthermore, detached eclipsing binaries have a special status because not only precise individual masses but also
physical dimensions can be obtained \citep{tor} and stellar evolution theory can be put to test, with the number 
of suitable objects counting in dozens.

Wide binaries with periods longer than 100 yr are even harder to characterize even if they are resolved in general
astrometric catalogs. Both astrometric and spectroscopic observations should be decades long to produce robust
orbital solutions and mass ratio estimates. Techniques have been proposed to statistically estimate or constrain
some of the essential parameters. For example, the relative velocity of components can be used to estimate the
eccentricity \citep{tok20}. Apart from the traditional position measurements, astrometric catalogs provide information
about the apparent acceleration of stars, which can be used to detect previously unknown binary systems from the
observed change of their proper motions \citep{wie} and, in some cases, to set constraints on the masses of unresolved
companions \citep{mak05}. Apparent angular accelerations were introduced into the basic astrometric model for the first
time in the Hipparcos main mission. As a detection tool for new binaries, they proved partially successful
as some of them could not be confirmed by high-resolution and spectroscopic follow-up observations, but more
sophisticated statistical analysis can produce more efficient selections of candidates \citep{fon}. Physical accelerations
can also be used to estimate component masses is several specific cases when combined with astrometric acceleration data,
epoch position measurements, and spectroscopic radial velocity observations \citep{bra}.

The goal of this paper is to demonstrate that accurate mass ratios can be derived for resolved long-period binary stars
from precision astrometry at two separate epochs, e.g., from the data in the Hipparcos \citep{esa} and Gaia EDR3
\citep{bro} catalogs, for potentially thousands of objects. No astrophysical assumptions or additional data are
needed for this method, and only the absolute positions are required from the first-epoch catalog. As a side product,
the proper motion of the system's barycenter is obtained, which can be used in kinematic studies of binary members
of nearby open clusters, for example. The main conditions of applicability are that the orbital period should be
much longer than the duration of each astrometric mission, which is about 4 years for the considered example, and the system
should be close enough to produce measurable changes in the apparent orbital motion of the components. The method is
presented in Section \ref{q.sec}. Its specific application is described on the example of a known nearby multiple
star HIP 473 in Section \ref{473.sec}. Realistic estimation of the resulting mass ratio error (uncertainty), taking
into account the full variance-covariance matrix of the input data, is essential, and a Monte Carlo-based technique
is described in Section \ref{unc.sec}. Section \ref{tri.sec} describes a generalization of this method to triple
systems, in which case mass ratios of two components can be directly computed, or approximately estimated using a
weighted generalized least-squares adjustment within the Jacobi coordinates paradigm. The prospects of
applying this new method to extensive surveys and space astrometry catalogs are discussed in Section \ref{con.sec},
as well as possible caveats related to the required absolute character of the utilized astrometric data.

\section{Mass ratios from astrometry}
\label{q.sec}
Let a binary star system with two components $i=1,2$ have position unit-vectors $\boldsymbol{r}_{ij}$ in two
astrometric catalogs, $j=1,2$, at epochs $t_1$ and $t_2$. If $\boldsymbol{r}_{0j}$ is the mean position of
the barycenter at these epochs, the vectors $\boldsymbol{\rho}_{ij}=\boldsymbol{r}_{ij}-\boldsymbol{r}_{0j}$ represent the angular separations
of the components, which are, to a very good approximation, equal to the projected orbital radii\footnote{
the small difference caused by the the curvature of the celestial sphere and the non-orthogonality
of these vectors to $\boldsymbol{r}_{0j}$ can be safely neglected.}. The projection onto the tangent plane preserves
the center of mass definition, so that
\eb
m_1\boldsymbol{\rho}_{1j}+m_2\boldsymbol{\rho}_{2j}=0,
\label{rho.eq}
\ee
where $m_i$ is the mass of the $i$th component. The barycenter's proper motion
\eb
\boldsymbol{\mu}_0=\frac{\boldsymbol{r}_{02}-\boldsymbol{r}_{01}}{\Delta t},
\ee
where $\Delta t$ is the epoch difference $t_2-t_1$, is assumed to be constant in time, which is also a
good approximation for realistic epoch differences. The projected orbital motion of component $i$ at epoch $j$
is
\eb
\boldsymbol{\psi}_{ij}=\boldsymbol{\mu}_{ij}-\boldsymbol{\mu}_0,
\ee
where $\boldsymbol{\mu}_{ij}$ is the observed proper motion. From differentiating Eq.~\ref{rho.eq} in time, we can see that
\eb
m_1\boldsymbol{\psi}_{1j}+m_2\boldsymbol{\psi}_{2j}=0.
\label{nu.eq}
\ee
Dividing this equation by $m_1$ and introducing $q=m_2/m_1$ obtains
\eb
\boldsymbol{\mu}_0=\frac{\boldsymbol{\mu}_{1j}+q\boldsymbol{\mu}_{2j}}{1+q}.
\label{mu0.eq}
\ee
Taking the difference of Eqs.~\ref{rho.eq} between the two epochs, dividing it by $m_1$, and substituting Eq.~\ref{mu0.eq}
obtains
\eb
\boldsymbol{r}_{12}-\boldsymbol{r}_{11}+q(\boldsymbol{r}_{22}-\boldsymbol{r}_{21})=(\boldsymbol{\mu}_{1j}+q\boldsymbol{\mu}_{2j})\Delta t.
\label{eq.eq}
\ee
All the terms in this equations are known from observation except $q$, which makes an overdetermined system
of up to four condition equations with one unknown. Unfortunately, the proper motions of visual binaries
in the Hipparcos catalog, which is used as the first-epoch catalog, are too unreliable to be used with
this method \citep{ma20}, so that only $j=2$ can be used, leaving two condition equations. Introducing the long-term
apparent motion of the component $\boldsymbol{\nu}_i=(\boldsymbol{r}_{i2}-\boldsymbol{r}_{i1})/\Delta t$, the exact solutions for $q$ can be written as
\eb
q=\frac{\boldsymbol{\mu}_{12}-\boldsymbol{\nu}_1}{\boldsymbol{\nu}_2-\boldsymbol{\mu}_{22}}.
\label{sol.eq}
\ee
Note that we have not made any assumptions about the masses of component in this derivation, and the estimated
$q$ can be negative. The vectors in the numerator and denominator are not parallel because of observational errors,
which effectively leads to two equations when their coordinate projections are considered. The two emerging
estimates of $q$ can be averaged with weights; however, as we will see in the example of HIP 473, one of the
projections can be practically useless when the orbital acceleration is almost aligned with the other coordinate
direction. It is simpler to use the ratio of the norms of the two calculated vectors:
\eb
q=\frac{\norm{\boldsymbol{\mu}_{12}-\boldsymbol{\nu}_1}}{\norm{\boldsymbol{\nu}_2-\boldsymbol{\mu}_{22}}}.
\label{nor.eq}
\ee
Alternatively, a more complex approach can be considered involving projections of the two observed vectors onto
a common unit vector, which minimizes the variance of $q$ from the combined contribution of the corresponding
error ellipses. The degree of misalignment, which is defined as
\eb\eta= \arccos{\left(\boldsymbol{v}_1\cdot \boldsymbol{v}_2/(\norm{\boldsymbol{v}_1} \norm{\boldsymbol{v}_2})\right)},
\label{eta.eq}
\ee
where $\boldsymbol{v}_1$ and $\boldsymbol{v}_2$ are the vectors in the numerator and denominator of Eq. \ref{sol.eq}, is a useful diagnostics of the reliability of estimation.

\section{Application to the wide binary system HIP 473}
\label{473.sec}

The long-period binary star HIP 473 = HD 38 = WDS 00057$+$4549 is one of the better studied systems due to its proximity and brightness.
It consists of two well-separated ($\rho=6.041\arcsec$) late-type dwarfs A and B of spectral types K6V and M0.5V
\citep{joy,kee,tam}. Earlier spectroscopic observations suggested twin component of the same spectral type M0 \citep{ano}.
\citet{kiy} used multiple photographic position measurements, as well as previously available radial velocity estimates
from \citep{tok90}, to produce a first reliable orbital solution for the inner pair, as well as preliminary estimates for the
orbit of a distant physical tertiary known as component F in the WDS. The latter has only a very short segment available
from astrometric observations due to its very long period. The method of {\it Apparent Motion Parameters} specially
designed by \citet{kis,kk} for such poorly constrained solutions was employed, which required the knowledge of the center
of mass for the AB pair. \citet{kiy} assumed equal masses for the A and B companions, which was a guess based on the
spectroscopic and photometric similarity of these stars \citep[cf.][]{luc,cve}. For the inner pair, 
they estimated a semimajor axis $a=6.21\pm 0.77\arcsec$,
a period $P=509.65\pm96.99$ yr, and an eccentricity $e=0.22\pm 0.04$. Their assumption about the relative masses of these
components allowed them to draw interesting conclusions about the relative inclination of the inner and outer orbits.
The estimated total mass $1.4\,M_{\sun}$, on the other hand, appears to be higher than what the spectral type would suggest.
The primary star HD 38 was already included in the first edition of the {\it Catalog of Chromospherically Active Binary Stars}
by \citet{str}. The signs of chromospheric activity in M dwarfs, such as H$_\alpha$ emission lines and detectable X-ray
radiation, are often related to rapid rotation fueled by a close orbiting companion. More recent radial velocity
measurements indicate that both A and B companions may be binary systems \citep{spe}, which would make the
entire system at least quintuple. Owing to its brightness and relatively wide separation, the AB pair has been
vigorously pursued with speckle and direct imaging CCD camera astrometry \citep{mas07,cve11,mas12,har} aiming at a better characterization of its orbit and
component masses. Despite this effort and many decades of observation, there is some ambiguity about the orbit.
A much shorter-period orbit with a higher
eccentricity was proposed by \citet{izm}: $P=370.7$ yr, $e=0.48$. Consequently, the total mass for this version is
substantially smaller, yielding a mass of $\sim 0.6\,M_{\sun}$ for each component, when equally divided. There is no
observational information about the mass ratio. Potentially, this parameter can be estimated from the gradient of
individual radial velocities, which requires a similarly persistent multi-year spectroscopic campaign. It can also
be obtained from Eq.~\ref{nor.eq} using publicly available astrometric data from the Hipparcos and Gaia EDR3 catalogs.

We begin with extracting the RA and Dec coordinates for both components at the corresponding mean epochs 1991.25
and 2016. The separation barely changed in the intervening 24.75 years, but the relative orientation did change
because of the orbital motion by approximately $10\degr$ in position angle. These coordinates are used to compute
the long-term proper motions vectors $\boldsymbol{\nu}_1$ and $\boldsymbol{\nu}_2$. They happen to differ from the short-term proper motions
taken directly from Gaia EDR3 by several mas yr$^{-1}$, which is a large multiple of their formal errors. The high 
signal-to-noise ratio hints at an accurate determination (\S\ref{unc.sec}). Indeed, the orbital signals for both
components exceeds 5 mas yr$^{-1}$ in the declination coordinate, which is much greater than their formal errors.
The computation of $q$ by Eq. \ref{nor.eq} is straightforward resulting in $q=0.996$. The measure of misalignment
$\eta$ (Eq. \ref{eta.eq}) caused by the astrometric errors is comfortably small at $2.5\degr$.

Binaries with near-twin components, such as the HIP 473 AB system, are more amenable to the
proposed astrometric estimation of $q$ because the observed signal is more equally distributed between
the components in Eq. \ref{nor.eq}. Numerous systems with small $q$ values are harder to process because the
signal in the numerator becomes too small compared to the expectation of observational error. Apart from
the misalignment parameter $\eta$, a useful quantity to filter out unreliable solutions is the 
empirical signal-to-noise
(SNR) parameter, which can be calculated as ${\rm SNR}=\sqrt{\boldsymbol{v}'_1\,\boldsymbol{C}^{-1}_1\,\boldsymbol{v}_1}$. On the other hand, low-mass companions are often much fainter than the primaries,
and the quality of Hipparcos astrometry quickly deteriorates for magnitudes approaching the sensitivity
limit $H_p\approx 11$ mag.

\section{Estimation of uncertainties}
\label{unc.sec}
The estimated mass ratio is the ratio of two random variables (Eq.~\ref{nor.eq}) derived from the astrometric data in two
separate catalogs. Given the complete variance-covariance matrices of the astrometric parameters, it is possible to
analytically estimate the resulting variance of $q$ in the large signal-to-noise approximation, because the
numerator and denominator are assumed to be statistically independent\footnote{In truth, the astrometric data for
nearby stars in a space mission astrometric catalog are always positively correlated because of the common attitude and
calibration unknowns in the condition equations. Since the global normal equations are not available, these correlations are
not accurately known, and are usually neglected.}. A more general and reliable approach, which is presented in this Section,
utilizes Monte Carlo simulations and confidence interval estimation. It is valid in the low signal-to-noise regime
too, providing an additional indication of problematic cases. The greatest weakness of both techniques is the underlying
assumption of Gaussian-distributed noise in the input data.

The five astrometric parameters for a regular solution are indexed 1 through 5 in this order: Right ascension coordinate,
declination coordinate, parallax, proper motion in right ascension, proper motion in declination. The associated covariance
matrix $\boldsymbol{C}$ is a $5\times 5$ symmetric matrix that can be reconstructed from the 5 formal standard errors and 10 correlation
coefficients given in the catalog. For the first-epoch catalog (Hipparcos), we only need the position covariances from the block
in the intersection of 1st and 2nd rows and 1st and 2nd columns, which is denoted as $(1,2;1,2)$ here. The second-epoch
catalog (Gaia) contributes three $2\times 2$ bocks, namely, the position covariance $(1,2;1,2)$, the proper motion covariance
$(4,5;4,5)$, and the off-diagonal position -- proper motion covariance $(1,2;4,5)$. The total covariance of vector $v_i$
is then
\eb
\begin{split}
\boldsymbol{Cov}[v_i]=&\boldsymbol{C}_{i1(1,2;1,2)}\Delta t^{-2}+\boldsymbol{C}_{i2(1,2;1,2)}\Delta t^{-2}+\\
&         \boldsymbol{C}_{i2(4,5;4,5)}-2 \boldsymbol{C}_{i2(1,2;4,5)}\Delta t^{-1}.
\label{cov.eq}
\end{split}
\ee

For the HIP 473 binary, I computed these covariances from the data in the {\it Double and Multiple System Annex} (DMSA)
of the Hipparcos catalog and the Gaia EDR3 catalog. DMSA does not provide the correlations $\boldsymbol{C}_{i1(1;2)}$, so I used
the correlation coefficient from the main catalog for the primary component (A), which is $-0.05$,
and assumed the same value for the B component. For each component, 2000 random $\cal{N}$$(0,1)$-distributed 2-vectors
were generated. If $\boldsymbol{p}$ is one of such vectors, a single randomly generated realization of $\boldsymbol{v}_i$ with Gaussian additive errors is
$\boldsymbol{v}_i+\boldsymbol{Cov}[v_i]^{1/2}\cdot \boldsymbol{p}$\footnote{The matrix square root of $\boldsymbol{x}$ is computed by functions {\tt sqrtm(x)} in MATLAB or
{\tt MatrixPower[x, 1/2]} in Mathematica.}. Each of these realizations results in a randomly dispersed $q$ by Eq.~\ref{nor.eq}.

\begin{figure}\centering
\includegraphics{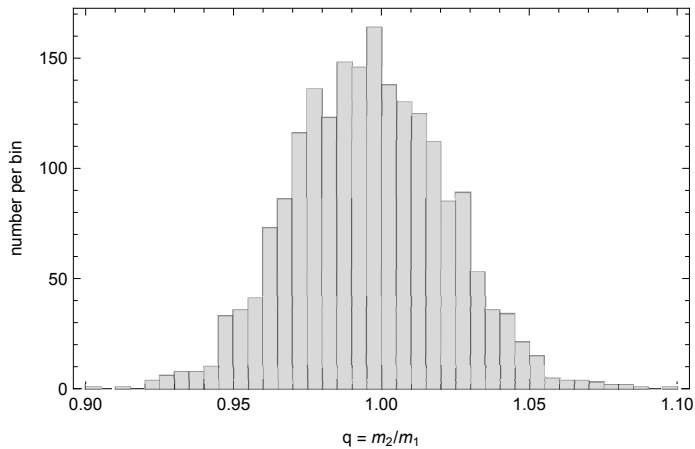}
\caption{Monte Carlo simulated distribution of the mass ratio $q$ for the binary star HIP 473 with nominal astrometric
parameters and variance-covariance matrices from the Hipparcos DMSA and Gaia EDR3. 2000 random trials were computed for
each of the two components.
\label{q.fig} }
\end{figure}

Fig. \ref{q.fig} shows the histogram of 2000 Monte Carlo estimates of $q$ for HIP 473 with astrometry from Hipparcos 
and Gaia EDR3. The sampled distribution is bell-shaped and fairly symmetric with a well-defined core. 
The standard deviation errors on the negative and positive sides can be estimated directly from the 
0.1573 and 0.8427 quantiles of the sampled distribution, which are not necessarily equal in absolute value. The result
for this binary star is $q=0.996^{+0.026}_{-0.026}$. Alternatively, we deduce from the 0.01 quantile
that $q>0.935$ with a 99\% confidence
given the data in Hipparcos and Gaia EDR3. The major part of the uncertainty comes from the position errors of the
Hipparcos catalog.

\section{Generalization for triple stars}
\label{tri.sec}
The orbital motion of isolated hierarchical triple star systems is long-term non-Keplerian, with the trajectories
subject to secular changes of sometimes drastic character, including flips of inner orbit orientation and reversal
of orbital spin \citep{nao}. On the time scales of one orbital revolution and shorter, however, the inner and outer
trajectories can be well approximated with Kepler's orbits of fixed elements. Within this approximation, the paradigm
of Jacobi coordinates simplifies the consideration of relative orbital motion. We consider a mathematical center of
mass of the inner pair components 1 and 2, which is located at $\boldsymbol{r}_0$. This point is involved in orbital motion with
the tertiary at $\boldsymbol{r}_3$ around the barycenter of the entire triple system at $\boldsymbol{r}_{\rm B}$ as if there is a physical
body of a $m_1+m_2$ mass. For the proper motions, we can write
\eb
m_3\,\bar{\boldsymbol{\mu}}_3+(m_1+m_2)\,\bar{\boldsymbol{\mu}}_0=0,
\ee
where $\bar{\boldsymbol{\mu}}_3$ and $\bar{\boldsymbol{\mu}}_0$ are the proper motion parts caused by the projected orbital motion, and index $j$
defining the epoch of observation is dropped for brevity. Like in a two-body problem, the actual or ``observed"
proper motions are
\begin{eqnarray}
\boldsymbol{\mu}_3&= \boldsymbol{\mu}_{\rm B}+\bar{\boldsymbol{\mu}}_3 \nonumber\\
\boldsymbol{\mu}_0&=\boldsymbol{\mu}_{\rm B}+\bar{\boldsymbol{\mu}}_0
\end{eqnarray}  
where index B denotes the current center of mass of the entire system. Within the Jacobi coordinates paradigm,
the observed components of the inner pair are
\begin{eqnarray}
\boldsymbol{\mu}_1&=\boldsymbol{\mu}_0+\bar{\boldsymbol{\mu}}_1 \nonumber\\
\boldsymbol{\mu}_2&=\boldsymbol{\mu}_0+\bar{\boldsymbol{\mu}}_2,
\end{eqnarray}  
with the orbital components still obeying the local inertial motion condition
\eb
m_1\bar{\boldsymbol{\mu}}_1+m_2\bar{\boldsymbol{\mu}}_2=0.
\ee
This leads to the general equation
\eb
m_1\boldsymbol{\mu}_1+m_2\boldsymbol{\mu}_2+m_3\boldsymbol{\mu}_3=(m_1+m_2+m_3)\boldsymbol{\mu}_{\rm B},
\label{mu3.eq}
\ee
despite the vectors connecting 0 with 3 and 1 with 2 being possibly not coplanar. This can also be derived by
differentiating the position equation in time:
\eb
m_1 \boldsymbol{r}_{1j}+m_2\boldsymbol{r}_{2j}+m_3\boldsymbol{r}_{3j}=(m_1+m_2+m_3)\boldsymbol{r}_{j{\rm B}}.
\ee
Taking the difference between epochs 1 and 2 results in
\eb
m_1 \boldsymbol{\nu}_1+m_2\boldsymbol{\nu}_2+m_3\boldsymbol{\nu}_3=(m_1+m_2+m_3)\boldsymbol{\mu}_{\rm B},
\label{nu3.eq}
\ee
where we assumed that the system's barycenter moves along a linear trajectory in space with a constant velocity.
Eliminating the unknown barycenter proper motion in Eqs. \ref{mu3.eq} and \ref{nu3.eq} leads to the final
condition equations
\eb
q_2(\boldsymbol{\nu}_2-\boldsymbol{\mu}_2)+q_3(\boldsymbol{\nu_3}-\boldsymbol{\mu}_3)=-(\boldsymbol{\nu}_1-\boldsymbol{\mu}_1).
\label{qq.eq}
\ee
There are two unknowns, $q_2=m_2/m_1$ and $q_3=m_3/m_1$, and two equations, so an explicit solution is available.
This would neglect the uncertainties of the vectors both in the condition and the right-hand part of the
equations. Alternatively, a generalized least-squares solution can be applied utilizing the known covariances.
This can be technically realized as a least-squares solution with a left-multiplied non-diagonal weight
matrix $\boldsymbol{\Omega}^{-1}$, which is the inverted sum of all three individual covariance matrices, $\boldsymbol{\Omega}=\boldsymbol{C}_1+\boldsymbol{C}_2+\boldsymbol{C}_3$,
each computed by Eq.~\ref{cov.eq}. The solution is then $[q_2,q_3]'=(\boldsymbol{X}'\,\boldsymbol{\Omega}^{-1}\boldsymbol{X})^{-1}\boldsymbol{X}'\,\boldsymbol{\Omega}^{-1}\boldsymbol{y}$,
where $\boldsymbol{X}$ is the design matrix and $\boldsymbol{y}$ the right-hand part of condition equations \ref{qq.eq}.
The main limitation of this method comes from the fact that the components of $\boldsymbol{\nu}_3-\boldsymbol{\mu}_3$ appear in the denominator
of the explicit solution. The orbital motion of tertiary stars may be so slow that only the nearest hierarchical
systems are suitable for this analysis. On the other hand, a weak signal for the tertiary allows one to ignore the
apparent acceleration of the inner pair due to its presence, and apply the algorithm for isolated binary systems.

\section{Discussion and future work}
\label{con.sec}

The proposed method of mass ratio estimation for resolved binary stars requires only precision astrometric positions
from two separate epochs and proper motions from one epoch. It is free of any astrophysical assumptions or models
about the physical parameters of the component stars. It should be emphasized, however, that the astrometric data
need to be {\it absolute}, not differential. In other words, the components' positions and proper motions should be
referenced to a well-established, non-rotating celestial reference frame. This brings up the issue of reference
frame consistency between the available astrometric solutions. The most obvious application is the Hipparcos-Gaia
pair, now separated by 24.75 years. The alignment of these important optical frames remains an open issue. 
Gaia DR2 and Hipparcos have a statistically significant misalignment, which emerges when Gaia positions of
brighter stars are transferred
to 1991.25 and compared with Hipparcos mean positions \citep{mb}. Based on archival VLBI astrometry of 26 selected
radio stars (which are optically bright), \citet{lin} concluded that the Gaia DR2 proper motion field includes a
rigid spin of approximately 0.1 mas yr$^{-1}$ for brighter stars only. Therefore, the global spin of the Gaia EDR3
frame was technically adjusted by a similar amount via a specific instrument calibration parameter for brighter stars
\citep[$G<13$ mag,][]{linast}. If the origin of this systematic error is correctly identified, this ad hoc correction
should remove the estimation bias of $q$. There remains a possibility that the discrepancy comes in part from a misalignment
of the Hipparcos frame, or Gaia EDR3 frame, or both, with respect to a common inertial reference frame. Even though
the systematic error is practically equal for the two components of a binary system, it comes with a different sign
with respect to the true orbital signal $\nu_i-\mu_i$, and the impact may be significant. For example, the estimated
misalignment of the Gaia frame by 1.4 mas results in a perturbation in the components' orbital signal of up to 57 $\mu$as
yr$^{-1}$, which can be critical for nearby long-period pairs such as HIP 473. Further progress and performance improvement
depends on our ability to evaluate and correct the remaining orientation misalignments and spins of the optical reference
frames.

Resolved triple systems can be processed as well providing both mass ratios for the
companions. Unresolved and unrecognized triples, on the other hand, is a threat to the proposed method
because the observed proper motions are perturbed by the orbital motion of the close binaries. The
effective perturbation is not always easy to estimate even for known single-lined spectroscopic
orbits as the photocenter effect is magnitude- and color-dependent. If the orbital period of the inner
pair is shorter than the mission length (about 4 years), the perturbation may be significantly averaged
out, so that the estimated mass ratio may still be of value. The inner orbital perturbation tends to
statistically increase the norm of the proper motion difference. If the primary (more massive component)
is a tight binary, this perturbation mostly increases the estimated $q$ sometimes resulting in values much exceeding 1. On the contrary, unresolved orbital motion of the secondary would decrease the estimated mass ratio
leading to an excess of systems with nearly zero $q$. The proposed SNR parameter is not helpful in such
cases, but the misalignment angle $\eta$ remains an efficient tool for vetting perturbed and unreliable
solution. Therefore, this method can be used to identify hidden hierarchical triples among resolved
double stars.

The method of astrometric mass ratio determination described in this paper requires absolute positions of components
at two epochs separated by a significant fraction of the orbital period, plus a proper motion from one of the epochs.
This points at the Hipparcos-Gaia combination as the most obvious source of data. Numerous differential observations
with speckle cameras, adaptive optics imaging, or long-base interferometry cannot be used, unfortunately, unless
they are ``absolutized" by direct or indirect reference to a well-established realization of the inertial
celestial frame. For example, HST images of fainter binaries can be reprocessed using the Gaia sources in the
field of view as astrometric reference. VLBI measurements of double radio sources are mostly obtained in the
phase-reference regime and are anchored to the nearby ICRF calibrators. Furthermore, this method can be trivially
generalized for three separate epochs of position astrometry without the need for precision proper motions. 

Accurate mass ratio estimation requires a sufficiently high signal-to-noise ratio. The main contribution to
the estimation uncertainty for the HIP 473 AB pair, used as an example in this paper, comes from the formal errors of
Hipparcos positions. This cannot be improved, and the application is currently limited to nearby systems
and binaries with orbital periods within a few hundred years. For a circular face-on orbit
with a period much longer than the epoch difference ($\Delta t<< P$), the total proper motion change
can be estimated as 
\eb
\Delta \mu \approx (2\pi)^2 M^\frac{1}{3}\varpi \Delta t P^{-\frac{4}{3}}.
\ee
We also note that binaries with near twins are favourable for this method because the $\Delta\mu$ signal
is more evenly distributed between the companions.
Among the
future plans, the proposed Gaia-NIR space astrometry mission holds the best promise \citep{hob,mca}. Having two
epochs of absolute astrometry at the Gaia's level of accuracy separated by 25--30 years will lead to characterization
of millions of fainter and more distant binaries.

\section*{Acknowledgments}

\end{document}